\documentclass[sigconf]{acmart}
\usepackage{multirow}
\usepackage{listings}
\usepackage{subcaption}
\DisableLigatures[-]{}
\copyrightyear{2026}
\acmYear{2026}
\setcopyright{cc}
\setcctype{by}
\acmConference[ICSE-NIER '26]{2026 IEEE/ACM 48th International Conference on Software Engineering}{April 12--18, 2026}{Rio de Janeiro, Brazil}
\acmBooktitle{2026 IEEE/ACM 48th International Conference on Software Engineering (ICSE-NIER '26), April 12--18, 2026, Rio de Janeiro, Brazil}
\acmPrice{}
\acmDOI{10.1145/3786582.3786825}
\acmISBN{979-8-4007-2425-1/2026/04}
\begin{document}
\title{Leveraging Design-Aware Context in Large Language Models for Code Comment Generation}

\author{Aritra Mitra}
\orcid{0000-0003-0248-1509}
\affiliation{%
\institution{Indian Institute of Technology Kharagpur}
\country{India}
}
\email{aritramitra2002@gmail.com}
\authornote{These authors contributed equally to the paper.}
\author{Srijoni Majumdar}
\orcid{0000-0003-3935-4087}
\affiliation{%
\institution{University of Leeds, UK}
\country{ }
}
\affiliation{%
\institution{Indian Institute of Technology Kharagpur}
\country{India}
}
\email{s.majumdar@leeds.ac.uk}
\authornotemark[1]
\author{Anamitra Mukhopadhyay}
\orcid{0009-0008-8852-0217}
\affiliation{%
\institution{Indian Institute of Technology Kharagpur}
\country{India}
}
\email{anamitra137@gmail.com}
\author{Partha Pratim Das}
\orcid{0000-0003-1435-6051}
\affiliation{%
\institution{Ashoka University}
\country{India}
}
\email{partha.das@ashoka.edu.in}
\author{Paul D Clough}
\orcid{0000-0003-1739-175X}
\affiliation{%
\institution{University of Sheffield, UK}
\country{ }
}
\affiliation{%
\institution{TPXimpact, UK}
\country{ }
}
\email{paul.clough@tpximpact.com}
\author{Partha Pratim Chakrabarti}
\orcid{0000-0002-3553-8834}
\affiliation{%
\institution{Indian Institute of Technology Kharagpur}
\country{India}
}
\email{ppchak@cse.iitkgp.ac.in}
\begin{abstract}
  Comments are very useful in the code development workflow. With increased access to and capability of tools, novice coders have been creating a significant amount of codebases. However, due to lack of standardization, their comments are often ineffective and can increase the time taken for subsequent maintenance. This study investigates the utility of using Large Language Models (LLMs) to generate more effective comments. The study focuses on the feasibility of utilizing design documents as a context for LLMs to generate more useful comments. Our results show that the usefulness of comments increases when using such contextual information.
\end{abstract}
\maketitle
\lstdefinestyle{codestyle}{
  language=C,
  basicstyle=\ttfamily\small,
  keywordstyle=\color{cyan}\bfseries,
  commentstyle=\color{teal}\itshape,
  stringstyle=\color{brown},
  backgroundcolor=\color{gray!10},
  frame=single,
  rulecolor=\color{gray!30},
  breaklines=true,
  xleftmargin=2pt,
  xrightmargin=2pt,
  aboveskip=2pt,
  belowskip=2pt
}
\lstset{style=codestyle}
\section{Introduction}
Natural Language Processing (NLP) and software engineering have both benefited greatly from the capabilities of Large Language Models (LLMs). These models are primarily trained to predict-the-next token. However, they can generalize and perform tasks, such as code summarization, test generation and code generation, especially after fine-tuning~\cite{yang2021comformercodecommentgeneration,8816757}. In this study, we focus on the task of generating code comments. Comments are the first layer of support for understanding codebases, offering file- and method-level context; whereas inline comments capture design choices~\cite{figl2025influencecodecommentsperceived,9610635}. Discrepancies between comments and code can help identify bugs and aid developers. Therefore, developers should write succinct, accurate, and up-to-date comments to guide future maintenance and use. However, comments are often neglected during development~\cite{4400153,10.1007/978-3-319-19243-7_10}. A good comment reduces maintenance cost while not bloating the codebase; ~\cite{981648,6171,chatterjee2025tool}; a bad comment provides little to no help and increases bloat~\cite{10.1016/j.jss.2022.111515}. This is especially important as typically the majority of the time spent by developers is understanding existing codebases ~\cite{7997917,chatterjee2023parallelc,paul2023efficiency}. \\
The comment generation task has previously been performed with code as context~\cite{ahmed2022fewshottrainingllmsprojectspecific,yang2021comformercodecommentgeneration,fried2023incodergenerativemodelcode,cui2022codeexpexplanatorycodedocument,venkatkrishna2023docgengeneratingdetailedparameter,shahbazi2021api2comimprovementautomaticallygenerated}, along with some additional information. However, in this work we use design documents as project-specific context. Design documents are commonly used to reference a codebase, as they explain the design choices, intended functionality, provide information on codebase architecture, system requirements, and contain links to any relevant sources~\cite{10.1145/1085313.1085331}. Hence, for a maintainer, design documents provide useful information while working on the codebase~\cite{dart1993case}. It is important, therefore, for developers to keep these documents up-to-date and accurate~\cite{dart1993case}. This study addresses the following research questions:
\begin{description}
  \item[RQ1:] Can design documents improve the quality of comments generated by LLMs?
  \item[RQ2:] Can Abstract Syntax Trees (ASTs)~\cite{sun2023abstractsyntaxtreeprogramming} be used as structural cues to generate better comments? 
\end{description}
We use Retrieval-Augmented Generation (RAG) ~\cite{lewis2021retrievalaugmentedgenerationknowledgeintensivenlp} to generate comments from the code where design documents are used as a source for retrieval. RAG models have been shown to provide substantial benefit over vanilla LLMs for knowledge-intensive tasks~\cite{gao2024retrievalaugmentedgenerationlargelanguage,lewis2021retrievalaugmentedgenerationknowledgeintensivenlp}. We have performed automated testing of the comments generated in this fashion using metrics, such as ROUGE~\cite{lin-2004-rouge}, BLEU~\cite{10.3115/1073083.1073135}, CodeBERTScore~\cite{cui2022codeexpexplanatorycodedocument} and GPTScore~\cite{fu2023gptscoreevaluatedesire}, as well as conducting a study with novice coders, providing them with bug-fixing and codebase enhancement tasks. We have also classified the LLM-generated comments by the kind of information they provide.

\begin{table}
\resizebox{\linewidth}{!}{
\begin{tabular}{|p{7cm}|p{8.75cm}|}
\hline
\textbf{Code} & \textbf{Comments} \\\hline\hline
\multirow{2}{*}{
\begin{minipage}[t]{\linewidth}
{\raggedright\ttfamily
\textcolor{cyan}{\bfseries void} check\_escaped(\textcolor{cyan}{\bfseries const char}* original,
\textcolor{cyan}{\bfseries const char}* escaped,
\textcolor{cyan}{\bfseries const bool} ensure\_ascii)\\
\{\\
\hspace*{2.5mm} std::stringstream ss;\\
\hspace*{2.5mm} json::serializer s(nlohmann::detail:: output\_adapter<\textcolor{cyan}{\bfseries char}>(ss), \textcolor{brown}{` '});\\
\hspace*{2.5mm} s.dump\_escaped(original, ensure\_ascii);\\
\hspace*{2.5mm} CHECK(ss.str() == escaped);\\
\}
}
\end{minipage}
}
&
\textbf{Comment generated: Few-Shot Prompting with Code} \newline
{\raggedright\ttfamily\itshape \textcolor{teal}{/** \newline * @brief Checks if a string is properly escaped as per JSON standards. \newline * @param original: The string to be escaped. \newline * @param escaped: The expected escaped string. \newline * @param ensure\_ascii: If true, ensures ASCII encoding. \newline */}
\par}
\\\cline{2-2}
 & \textbf{Comment generated: RAG with Design Document [API Documentation]} \newline
{\raggedright\ttfamily\itshape \textcolor{teal}{/** \newline ...\newline * Use Case: Test in json/test.\newline * Method used: \newline * - dump\_escaped, defined at json.hpp:8514\newline */}
\par}
\\\hline
\multirow{2}{*}{
\begin{minipage}[t]{\linewidth}
{\raggedright\ttfamily
\textcolor{cyan}{\bfseries static int} timer\_less\_than(\textcolor{cyan}{\bfseries const struct} heap\_node* ha, \textcolor{cyan}{\bfseries const struct} heap\_node* hb) \{ \\\hspace*{2.5mm} \textcolor{cyan}{\bfseries const} uv\_timer\_t* a, b; \\\hspace*{2.5mm} a = container\_of(ha, uv\_timer\_t, node.heap); \\\hspace*{2.5mm} b = container\_of(hb, uv\_timer\_t, node.heap); \\\hspace*{2.5mm} if (a->timeout < b->timeout)\\\hspace*{5mm} \textcolor{cyan}{\bfseries return} 1;\\\hspace*{2.5mm} if (b->timeout < a->timeout)\\\hspace*{5mm} \textcolor{cyan}{\bfseries return} 0;\\\hspace*{2.5mm} \textcolor{cyan}{\bfseries return} a->start\_id < b->start\_id; \\\} 
\par}
\end{minipage}
}
&
\textbf{Comment generated: Few-Shot Prompting with Code} \newline
{\raggedright\ttfamily\textcolor{teal}{/*\newline * Heap comparator function for timers.\newline *Primary: timeout, Secondary: start\_id for tie-breaking\newline * Prevents starvation of same-timeout timers\newline */}
\par}
\\\cline{2-2}
 & \textbf{Comment generated: RAG with Design Document [Implementation Document]} \newline
{\raggedright\ttfamily\itshape \textcolor{teal}{/*----------------------------------*\newline *  static int timer\_less\_than(const heap\_node* a, const heap\_node* b) \newline ...\newline *  LIMITATION: relies on start\_id never wrapping.\newline *----------------------------------*/}
\par}
\\\hline
\end{tabular}
}
\caption{Comparing comments generated with and without design documents. The comment generated with the implementation document detects an exception, and the comment generated with the API documentation locates a definition.}
\label{tab:gen_com_comp}
\end{table}
We evaluated four generation setups: (i) Few-shot prompting with 30 (at least 10 positive and 10 negative) examples, (ii) Few-shot prompting with condensed abstract syntax trees (ASTs) as added context, (iii) Few-shot prompting with RAG on the design document, and (iv) Combining ASTs with using RAG on the design document. In the cases with the design document, we have found an increased number of navigational comments: linking to other parts of the code; algorithmic comments: explaining the workings of the code; and exception-based comments: showing edge-cases of the code. This can be seen from the examples in Table \ref{tab:gen_com_comp} where LLM-generated comments with design documents perform better than comments without them. We have seen a 35\% decrease in bug-fixing time for LLM generated comments when the design document is used. The results for the codebase enhancement task are similar, except the decrease in time reduces to 28\%. This can be explained by not affecting the time taken to write the code; rather decreases come from easier understanding of code. To the best of our knowledge, this is the first systematic study of comment generation in C using an LLM architecture, with the following major contributions: \begin{itemize}
    \item A pipeline with the source code, ASTs, design documents, and RAG to support scalable and informative LLM-based comment generation for C and C++.
    \item An assessment framework for C/C++ comments based on usefulness, combining human and LLM judgments.
    \item Using design documents as context cues to gather project-specific information, and abstract syntax trees to gather system-specific information.
\end{itemize}
\section{Related Work}
The previous works in this line have dealt with evaluating~\cite{fan2024evaluatingqualitycodecomments} and generating~\cite{ahmed2022fewshottrainingllmsprojectspecific,yang2021comformercodecommentgeneration} summary comments using the code as context, albeit on different datasets~\cite{10.1007/s10664-019-09730-9,dong-lapata-2016-language,lu2021codexgluemachinelearningbenchmark}. Moreover, explanatory comment generation has been performed, using similar contexts~\cite{fried2023incodergenerativemodelcode,cui2022codeexpexplanatorycodedocument,venkatkrishna2023docgengeneratingdetailedparameter} on several datasets~\cite{lu2021codexgluemachinelearningbenchmark,8816757,cui2022codeexpexplanatorycodedocument}. However, these models have limited context lengths, which is often a limitation for large repositories. Hence, we focus on design documents, which encompass the design information regarding the repositories. Ahmed \textit{et. al.} ~\cite{ahmed2022fewshottrainingllmsprojectspecific} have used few-shot prompting~\cite{DBLP:journals/corr/abs-2005-14165} to show the necessity of project-specific information in comment generation. Shahbazi \textit{et.al.} ~\cite{shahbazi2021api2comimprovementautomaticallygenerated} have used API documentation, with RNN-based models instead of LLMs. We also focus on ASTs, as they provide structural information of the code. We also see that metrics like BLEU and ROUGE penalize verbose comments, including docstrings. Therefore, we use metrics such as BERTScore~\cite{zhang2020bertscoreevaluatingtextgeneration}. We also use human assessment as our metric, utilizing common tasks for a developer such as bug-fixing and code enhancement.
\section{Methods}
We have used the 27 comment categories~\cite{majumdar2022automated,fan2024evaluatingqualitycodecomments} classified by Majumdar \textit{et. al.} ~\cite{majumdar2022automated,majumdar2025comprehending}, focusing on the 8 categories for useful comments, as our gold standard. Table \ref{tab:param} shows an example of each class. We have also classified the design documents into several classes, according to ~\cite{cern1998introdoc}, as shown in Table \ref{tab:docdesc}. We have noted the frequency of occurrence of each type of document mined from 50 GitHub repositories.
\begin{table}
\resizebox{\linewidth}{!}{
\begin{tabular}{|p{12.3cm} c|}
\hline
\textbf{Comment Categories with Description and Examples} & \textbf{\%useful} \\
\hline\hline
{\bf Consistency:} The concepts in the comment correctly talk about the code \newline 
\lstinline|// using calloc to create 0-initialization| \newline
\lstinline|int *p=(int*)calloc(N, sizeof(int));| & 73.26 \\\hline
{\bf Algorithmic Details:} Explains working summary \newline   
\lstinline|/* This function tests the dump_escaped method of the json::serializer class to ensure that strings are properly escaped as per JSON standards. */| \newline
\lstinline|void check_escaped(const char* original, const char* escaped, const bool ensure_ascii);| & 65.56 \\\hline
{\bf Links:} The comment links the function to the files where it is defined or the header \newline
\lstinline|// dump_escaped defined in json.hpp:8514| \newline
\lstinline|s.dump_escaped(original, ensure_ascii);| & 36.88 \\\hline
{\bf Domain-Specific Context:} The comment links the program  with application domain concepts \newline 
\lstinline|/** Changes L-D orientation of amino acid by mirroring the bonds */| \newline
\lstinline|void _flip_amino_acid(amino *A);| & 35.31 \\\hline
{\bf Possible Exceptions:} The comment lists potential bugs that can arise due to change in parameter \newline  
\lstinline|// no bounds checking on addr.| \newline
\lstinline|int oapv_bsw_write_direct(void *addr, uint32_t val, int len);| & 30.66 \\\hline
{\bf Irrelevance:} The comment explains concepts already easily understandable from code \newline 
\lstinline|// Returns number of threads needed| \newline
\lstinline|int num_threads(int num_processes, char allocation_type, task* processes);| & 10.26 \\\hline
{\bf Alternative Solutions:} Provides alternate solutions to solve bugs, improve complexity \newline   
\lstinline|/* Alternative: use memcpy when bytes>=4. */| \newline
\lstinline|static int bsw_flush(oapv_bs_t *bs, int bytes);| & 9.14 \\\hline
{\bf Complexity:} Provides analysis on time and space complexity and also reasons for it \newline   
\lstinline|// O(k log n) for k expired timers| \newline
\lstinline|void uv__run_timers(uv_loop_t* loop);| & 8.19 \\\hline
\end{tabular}
}
\caption{The eight comment categories generally found in useful comments, from ~\cite{majumdar2022automated}.} \label{tab:param}
\end{table}
\begin{table}
\resizebox{\linewidth}{!}{
\begin{tabular}{|p{3cm}|p{10cm}|r|}
\hline
\textbf{Document Type} & \textbf{Purpose / Description} & \textbf{Frequency}\%\\
\hline
Requirements documents & Define functional, design, and operational requirements with unique ID, clear description, and attributes (priority, examples, acceptance criteria) & 24 \\
\hline
Architecture documents & Describe components and interfaces, show graphical overview and detailed text, specify functionality, decomposition, data exchange rules for modularity & 62 \\
\hline
Detailed Design & Expand architecture into fine-grained details. Use modeling methodologies (UML, OMT, Booch). Include use cases, data/object models, and behavior models & 16 \\
\hline
Implementation documents & Contain code and related artifacts, follow design choices, explain complex logic and reasoning, not trivial details & 72 \\
\hline
Test documents & Ensure product correctness. Include test plan (strategy), test cases (setup and verification), and test reports (execution results) & 48 \\
\hline
Project management documents & Plan and track the project. Include project plan, scheduling/tracking documents (Gantt, Milestone Trend charts), and regular team/customer reports & 28 \\
\hline
Configuration management documents & Describe configuration items of all project phases. Define repository structure, rules for modification, versioning, and releases & 64 \\
\hline
Project infrastructure documents & Guide project participants. Include conventions, tools, templates, reporting procedures, configuration system use, and role-specific instructions & 22 \\
\hline
User software documentation & For end-users: manuals, error messages, online help, websites, tutorials. Must be user-centered, task-oriented, concise, and tested with users & 96 \\
\hline
\end{tabular}
}
\caption{Summary of project and user documentation types with their purposes, from ~\cite{cern1998introdoc}.}
\label{tab:docdesc}
\end{table}
\\
Whilst choosing LLM models, we considered multiple factors: (i) We preferred reasoning models for our task, as they generated more meaningful comments, (ii) We wanted a large context length for the model to enable us to pass the entire source file as context, and (iii) We considered LLMs that specialize in code-specific tasks. After consideration, we selected \texttt{OpenAI o3}, \texttt{OpenAI o4-mini}~\cite{o4-mini}, \texttt{Codestral 25.01}~\cite{codestral}, \texttt{DeepSeek-R1}~\cite{deepseek-r1} and \texttt{GPT-4o}~\cite{GPT-4o} as the 5 models for our experiments. \\
We have also used the AST of the code from which we generate comments. However, ASTs are generated in compilation, and preprocessor macros can change them depending on the system. For this, we insert a command into the makefile to generate the AST and then build the code. Thus, we can get the AST that perfectly represents the code. We get the \texttt{CFLAGS} and \texttt{CPPFLAGS} from the makefile, and use \texttt{clang} with the flags to generate the ASTs for the codes. We use the design documents as data sources for RAG. We label design documents by their type ~\cite{cern1998introdoc}, and use all of them as the source. We provide the file with all of its headers expanded, and comments removed. The persona assumed by the LLM is a novice software developer with the task of code maintenance~\cite{salewski2023incontextimpersonationrevealslarge}.
\section{Experiment}
\subsection{Dataset Creation}
We chose GitHub repositories of libraries in C and C++ with high number of commits, stars and forks. We have chosen 9 repositories falling under these criteria. We have scraped the source codes and header files from these libraries, and created our dataset. The libraries we picked are as shown in Table \ref{tab:dataset}.
\begin{table}[!htbp]
    \resizebox{\linewidth}{!}{
    \begin{tabular}{|lcccc|}
    \hline
        Library & Language & Number of Code-Comment Pairs & Number of Source Files & Size of Design Context\\\hline\hline
        libPNG & C & 575 & 24 & 233KB\\
        postgresql.pthreads & C & 949 & 108 & 1.4MB\\
        libcurl & C & 2937 & 988 & 7.2MB\\
        libuv & C & 491 & 127 & 431KB\\
        libarchive & C & 2792 & 124 & 3.1MB\\
        openapv & C & 327 & 45 & 296KB\\
        spdlog & C++ & 239 & 152 & 351KB\\
        json & C++ & 3124 & 496 & 6.5MB\\
        taskflow & C++ & 684 & 53 & 2.8MB\\ \hline
    \end{tabular}
    }
    \caption{Characteristics of the dataset.}\label{tab:dataset}
\end{table}
\subsection{Prompt Engineering}
We have used several techniques to reduce potential biases introduced by prompting. (i) For \emph{exemplar bias}, we use a uniform distribution of good and bad comments in our prompts, so that the LLM can distinguish good from bad and create a good comment. (ii) For \emph{positional bias}, we have seen that changing the order of the examples showed a CodeBERTScore similarity of $\ge 0.95$ in the generated comments for 71\% of the cases. (iii) For \emph{format bias}~\cite{xu2024carepromptbiasinvestigating}, we have seen that changing the wording also showed a CodeBERTScore similarity of $\ge 0.95$ in the generated comments for 70\% of the cases. (iv) For \emph{knowledge bias}~\cite{xu2024carepromptbiasinvestigating}, we find that decreasing the context length only works for shorter codes, else the context limit is reached. However, the generated comments have a CodeBERTScore similarity of $\ge 0.95$ in 78\% of the cases.
\subsection{Participants and Tasks}
A total of 22 undergraduate computer science students were chosen for the experiments. While they had solid theoretical knowledge, they had limited practical experience with codebases or libraries, making them novice maintainers by industry standards. Each participant received 22 code samples: 1 without comments, 1 with original comments, and 20 with comments generated by LLMs (one for each LLM and context). 22 versions of each code were created, so every participant viewed only one version of each file. The methodology was as follows: (i) Generate comments with LLMs for each file using the appropriate context; (ii.a) For bug-fixing, introduce a runtime bug without changing major parts of the code, which the compiler does not detect; (ii.b) For enhancements in the codebase, remove a utility function but keep the related comment; (iii) Ask participants to fix the file; and (iv) Record the time taken to complete for quantitative analysis. Comments were generated before inserting bugs, creating intentional inconsistencies between code and comments helping to find bugs. All experiments were conducted on Intel x86-64 systems with Ubuntu 22.04 for consistency. Participants did not have access to LLMs, but could use the internet.
\section{Evaluation}
We use two approaches for evaluating the usefulness of generated comments, based on automated metrics and human experiments.
\subsection{Similarity with Human Annotations}
The first method of evaluation comes from the similarity of the LLM-generated comments with a known set of useful comments. We use the labeled dataset of comments from ~\cite{majumdar2022automated}, which is labeled as useful/useless by industry professionals, as the ground truth. \\
We choose CodeBERTScore ~\cite{cui2022codeexpexplanatorycodedocument}, GPTScore ~\cite{fu2023gptscoreevaluatedesire}, ROUGE ~\cite{lin-2004-rouge} and BLEU ~\cite{10.3115/1073083.1073135} as our metrics. We have chosen CodeBERTScore instead of BERTScore ~\cite{zhang2020bertscoreevaluatingtextgeneration}, because we want similarity in a code-specific context.
\begin{table*}
    \resizebox{\textwidth}{!}{
    \begin{tabular}{|c|cccc|cccc|cccc|cccc|}
        \hline
        Context & \multicolumn{4}{c|}{Code} & \multicolumn{4}{c|}{Code + AST} & \multicolumn{4}{c|}{Code + Design Doc} & \multicolumn{4}{c|}{Code + AST + Design Doc} \\
        Model & ROUGE-L & BLEU-4 & CodeBERTScore & GPTScore & ROUGE-L & BLEU-4 & CodeBERTScore & GPTScore & ROUGE-L & BLEU-4 & CodeBERTScore & GPTScore & ROUGE-L & BLEU-4 & CodeBERTScore & GPTScore \\
        \hline\hline
        \texttt{OpenAI o3} & 0.32 & 0.14 & \textbf{0.71} & \textbf{0.75} & \textbf{0.16} & 0.09 & \textbf{0.55} & \textbf{0.58} & 0.38 & 0.22 & \textbf{0.81} & \textbf{0.83} & \textbf{0.27} & 0.11 & \textbf{0.55} & \textbf{0.60} \\
        \hline
        \texttt{OpenAI o4-mini} & 0.31 & 0.14 & 0.69 & 0.70 & 0.16 & \textbf{0.09} & 0.52 & 0.50 & \textbf{0.39} & \textbf{0.24} & 0.79 & 0.79 & 0.23 & 0.10 & 0.52 & 0.54\\
        \hline
        \texttt{Codestral 25.01} & 0.27 & 0.12 & 0.67 & 0.70 & 0.15 & 0.08 & 0.49 & 0.47 & 0.34 & 0.18 & 0.73 & 0.70 & 0.21 & \textbf{0.12} & 0.46 & 0.46 \\
        \hline
        \texttt{DeepSeek-R1} & 0.29 & 0.10 & 0.66 & 0.70 & 0.14 & 0.06 & 0.46 & 0.51 & 0.36 & 0.20 & 0.69 & 0.73 & 0.19 & 0.10 & 0.52 & 0.50  \\
        \hline
        \texttt{GPT-4o} & \textbf{0.34} & \textbf{0.15} & 0.63 & 0.69 & 0.12 & 0.05 & 0.40 & 0.42 & 0.31 & 0.14 & 0.69 & 0.65 & 0.13 & 0.05 & 0.42 & 0.41 \\
        \hline
    \end{tabular}
    }
    \caption{Similarity of Human-annotated and LLM-generated comments in various contexts using automatic metrics.}
    \label{tab:similarity_metrics}
\end{table*}
From Table \ref{tab:similarity_metrics}, we see that performance drops sharply when ASTs are included, as their size crowds out useful context and lowers comment quality. Additionally, BLEU and ROUGE provide poor results because LLM-generated comments tend to be more verbose and detailed than those generated by humans, which these metrics penalize. Embedding-based metrics do not have this problem. Hence, we can focus on CodeBERTScore ~\cite{cui2022codeexpexplanatorycodedocument} and GPTScore ~\cite{fu2023gptscoreevaluatedesire}.
\subsection{Human Experiments}
We conducted two experiments: one for bug-fixing and one for enhancing new functionalities. We choose these because typically developers spend their majority of time on these tasks \cite{rollbar2021,linearb2024}. \\
\begin{table*}
    \begin{subtable}[t]{0.48\textwidth}
    \resizebox{\linewidth}{!}{
    \begin{tabular}{|l|cccc|}
        \hline
        Model & \multicolumn{4}{c|}{Time Taken (Average, minutes)} \\
        Context & Code & Code + AST & Code + Design Doc & Code + AST + Design Doc \\\hline\hline
        \texttt{OpenAI o3} & 19.77 & 24.68 & 17.14* & 24\\
        \texttt{OpenAI o4-mini} & 22.18 & 27.18 & 18.27* & 25.27\\
        \texttt{Codestral 25.01} & 25.86 & 30.82 & 22.68* & 29.73\\
        \texttt{DeepSeek-R1} & 25.45 & 30.5 & 22.77* & 29.73\\
        \texttt{GPT-4o} & 27.27 & 32.32 & 26.26* & 31.64\\
        \hline
        Original & \multicolumn{4}{c|}{26.36} \\
        No Comments & \multicolumn{4}{c|}{34.36} \\
        \hline
    \end{tabular}
    }
    \caption{Bug-fixing}
    \label{tab:debug_time}
    \end{subtable}
    \hspace{\fill}
    \begin{subtable}[t]{0.48\textwidth}
    \resizebox{\linewidth}{!}{
    \begin{tabular}{|l|cccc|}
        \hline
        Model & \multicolumn{4}{c|}{Time Taken (Average, minutes)}\\
        Context & Code & Code + AST & Code + Design Doc & Code + AST + Design Doc\\\hline\hline
        \texttt{OpenAI o3} & 28.32 & 33.55 & 25* & 32.36\\
        \texttt{OpenAI o4-mini} & 30.82 & 36.09 & 25.95* & 33.36\\
        \texttt{Codestral 25.01} & 33.36 & 38.64 & 29.82* & 37.23\\
        \texttt{DeepSeek-R1} & 33.41 & 38.64 & 30.64* & 38.09\\
        \texttt{GPT-4o} & 35.14 & 40.41 & 32.68* & 40.09\\
        \hline
        Original & \multicolumn{4}{c|}{35.05}\\
        No Comments & \multicolumn{4}{c|}{43.32}\\
        \hline
    \end{tabular}
    }
    \caption{Enhancements in a codebase}
    \label{tab:imple_time}
    \end{subtable}
  \caption{Time taken to perform the maintenance tasks. Values marked with * imply the context of Code + Design Doc performs better than the contexts of Code, Code + AST, Code + AST + Design Doc with statistical significance from two-tailed $\chi^2$-test~\cite{Pearson01071900}.}
\end{table*}
{\setlength{\abovecaptionskip}{-1pt}
\begin{figure*}
  \begin{subfigure}[b]{.5\linewidth}
    \includegraphics[width=.99\textwidth]{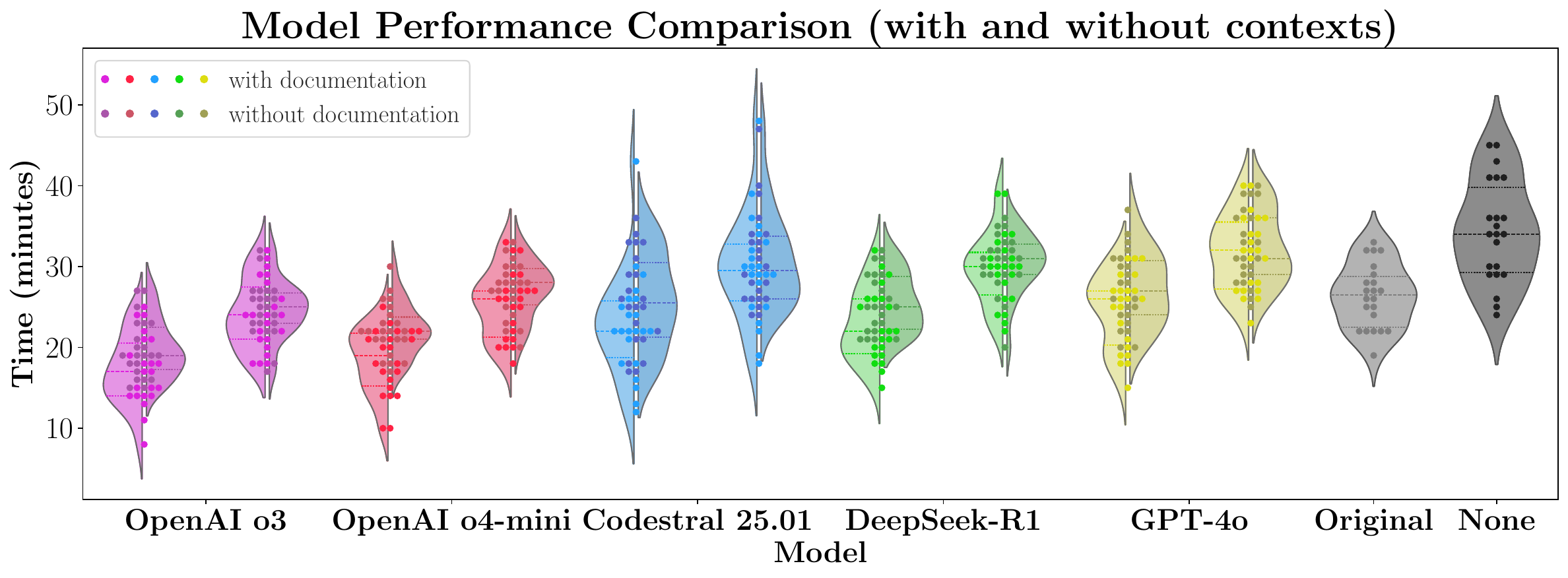}
    \caption{Time taken for bug-fixing.}\label{fig:Perf_d}
  \end{subfigure}%
  \begin{subfigure}[b]{.5\linewidth}
    \includegraphics[width=.99\textwidth]{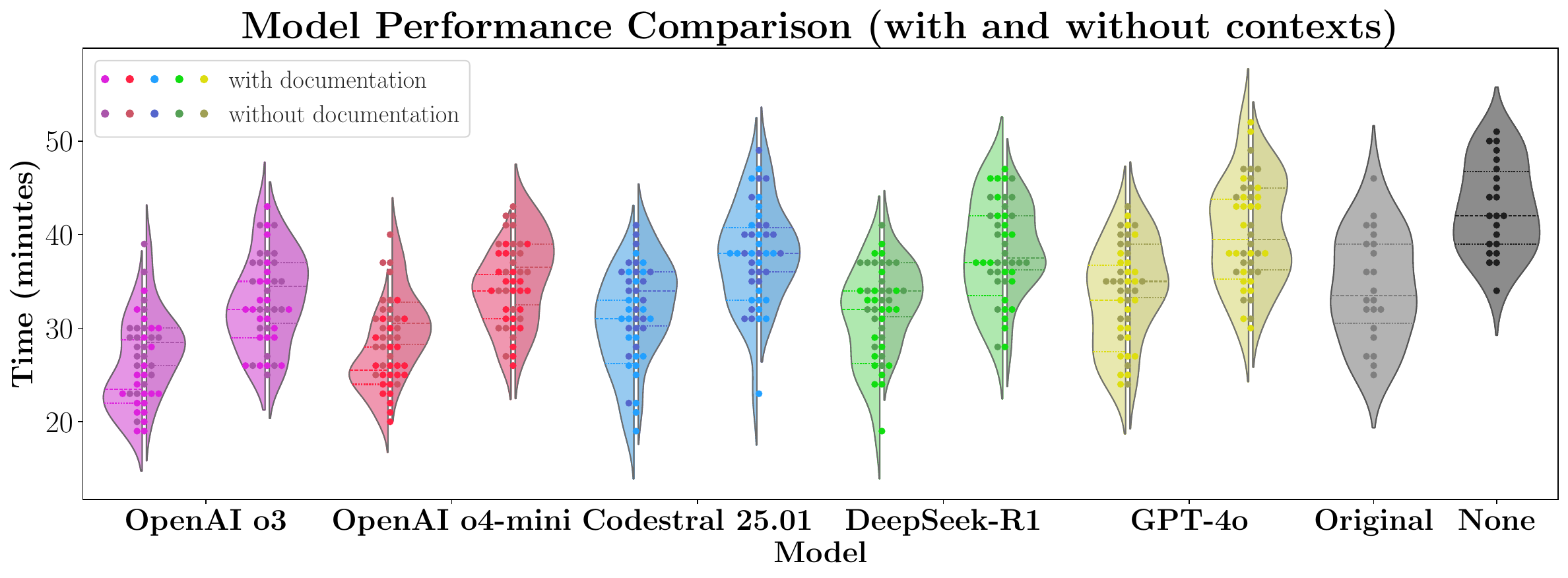}
    \caption{Time taken to enhance the codebase.}\label{fig:Perf_i}
  \end{subfigure}%
  \caption{Brighter points are with the design documents in context, and dimmer points are without them. For every LLM, the violins from left to right depict the context: (i) Code + Design Doc, (ii) Code, (iii) Code + AST + Design Doc, (iv) Code + AST.}
  \Description[Time taken to bugfix and enhance the codebase.]{The left image shows time taken for bugfixing, and the right image shows time taken to enhance the codebase. It shows that ASTs reduce performance, while design documents improve it, and that OpenAI o3 performs best among different LLMs.}
  \label{fig:Perf}
\end{figure*}
}\emph{\textbf{Bug-fixing.}} From Figure \ref{fig:Perf_d} and Table \ref{tab:debug_time}, we can clearly see that \texttt{OpenAI o3} significantly decreases the bug-fixing time. Also, comment-less code took much longer to fix, as there was no information on the bug. This shows that comments are useful for a maintainer, and that LLMs can generate comments which can be more useful for novice coders. However, we can also see that ASTs caused more harm than good as context and increased time taken. \\
\emph{\textbf{Enhancements in the codebase.}} We see a less pronounced, but similar effect, from design documents (Table \ref{tab:imple_time}). This occurs because comments reduce the time to understand and write code. Also, Figure \ref{fig:Perf_i} shows the effect of design documents and ASTs as context. While design documents hasten understanding, ASTs hinder it and make the task longer to complete.
\section{Discussion}
We have seen that with the documents already available in the repositories, as shown in Table \ref{tab:docdesc}, we can generate comments with a lightweight RAG-based implementation, which can facilitate better bug-fixing and codebase enhancement. Through this effective comment generation, we can move in a new direction, where we can develop a holistic view of the complete database through the comment, reverse engineering the design. We can also state that (\textbf{RQ1}) design documents can be used as project-specific context to generate better explanatory comments, helping novice coders; (\textbf{RQ2}) ASTs overwhelm LLMs with too much information, causing them to generate poorer comments. \\
\textbf{\emph{Threats to validity.}} There are several important points to note regarding comment generation using LLMs, which we discuss here.\\
\emph{Completeness.} When a larger piece of code is input to an LLM, parts of it are often omitted in the output. Moreover, this issue differs for different LLMs. \texttt{GPT-4o}, \texttt{OpenAI o4-mini}, \texttt{OpenAI o3} remove macros while keeping the core code intact; while \texttt{DeepSeek-R1} removes function bodies and replaces with \texttt{\{...\}}. \texttt{DeepSeek-R1} and \texttt{Codestral 25.01} also omit whole functions. To quantify this issue, we define \emph{Completeness Ratio} as the ratio of the sizes of the generated file and the original file, removing all comments and whitespace.
\begin{figure}
    \includegraphics[width=\linewidth]{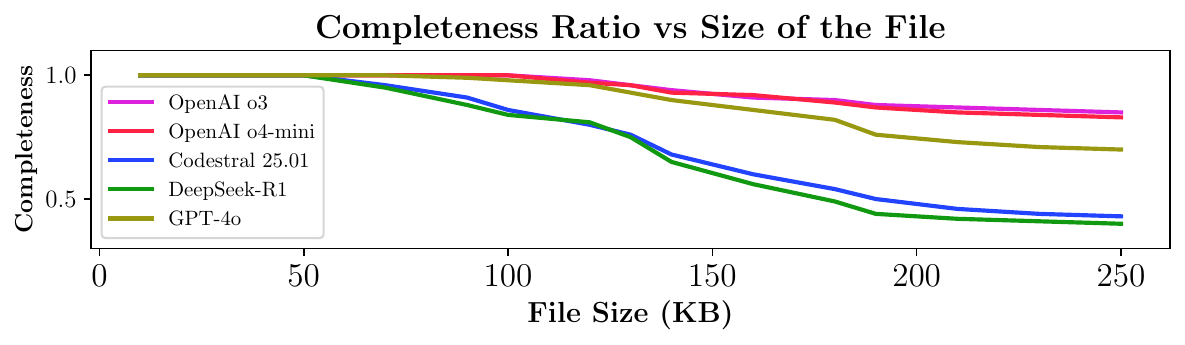}
    \caption{Decrease in completeness with increasing file size, showing deteriorating performance of the LLMs.}
    \label{fig:completeness}
    \Description[Completeness Ratio with respect to File Size]{Shows decrease in completeness with increasing file size, showing deteriorating performance of the LLMs.}
\end{figure}
Figure \ref{fig:completeness} shows a steady decline in the completeness ratio. Completeness is worst for \texttt{Codestral 25.01} and \texttt{DeepSeek-R1}, then in \texttt{GPT-4o}, then \texttt{OpenAI o3} and \texttt{o4-mini}, which fair similarly.
\emph{Context Size.} The context lengths of the LLM models are often smaller than a large file, so the comments for that file must be generated through several passes. This creates consistency issues: even though the coding guidelines mention the commenting style, comments of a previous function are important to comments further down. This nuance is lost when information from the previous function is not accessible while generating further comments.
\section{Conclusion}
We propose the use of design documents as a project-specific context for explanatory comment generation, using a RAG-based approach. We show the comments generated provide significant help to developers in their most frequent tasks. We also see the effects and caveats of different LLMs, with \texttt{OpenAI o3} performing the best.
\section{Future Plans}
Here, we focused on generating comments for novice developers using LLMs, using RAG with design documents as additional context. In the future, we plan to extend this study by changing the target demographic. We also aim to integrate the pipeline with existing IDEs, gaining IDE-specific information. Also, we can reduce computations by using smaller, but equally capable, models (possibly finetuned for our purpose). 
\section*{Acknowledgment}
We thank the developers at TPXimpact and Siemens EDA for their continuous support in annotating the data since 2021, and Professor Evangelos Pournaras for his guidance in writing niche papers.
\bibliographystyle{ACM-Reference-Format}

\bibliography{Bibliography}

@article{majumdar2025comprehending,
  title={Comprehending C codes with LLMs: Effective comment generation through retrieval and reasoning},
  author={Majumdar, Srijoni and Deshpande, Adwita and Das, Partha Pratim and Chakrabarti, Partha Pratim},
  journal={Pattern Recognition Letters},
  year={2025},
  publisher={Elsevier}
}

@inproceedings{paul2023efficiency,
  title={Efficiency of Large Language Models to scale up Ground Truth: Overview of the IRSE Track at Forum for Information Retrieval 2023},
  author={Paul, Soumen and Majumdar, Srijoni and Bandyopadhyay, Ayan and Dave, Bhargav and Chattopadhyay, Samiran and Das, Partha and Clough, Paul D and Majumder, Prasenjit},
  booktitle={Proceedings of the 15th Annual Meeting of the Forum for Information Retrieval Evaluation},
  pages={16--18},
  year={2023}
}

@article{chatterjee2025tool,
  title={Tool assisted Agile approach for legacy application migration},
  author={Chatterjee, Nachiketa and Majumdar, Srijoni and Das, Partha Pratim and Chakrabarti, Amlan},
  journal={International Journal of System Assurance Engineering and Management},
  volume={16},
  number={9},
  pages={3002--3017},
  year={2025},
  publisher={Springer}
}

@misc{GPT-4o,
  author       = {OpenAI},
  title       = {{Hello GPT-4o}},
  year         = {2024},
  url          = {https://openai.com/index/hello-gpt-4o/},
  note         = {Large Language Model.}
}

@misc{o4-mini,
  author       = {OpenAI},
  title       = {{Introducing OpenAI o3 and o4-mini}},
  year         = {2025},
  url          = {https://openai.com/index/introducing-o3-and-o4-mini/},
  note         = {Large Language Model.},
}

@misc{deepseek-r1,
  author  = {DeepSeek Research},
  title  = {{DeepSeek-R1: Incentivizing Reasoning Capability in LLMs via Reinforcement Learning}},
  howpublished = {\url{https://github.com/deepseek-ai/DeepSeek-R1/blob/main/DeepSeek_R1.pdf}},
  year    = {2025},
}

@misc{codestral,
  author       = {Mistral AI Team},
  title       = {{Codestral 25.01}},
  year         = {2025},
  url          = {https://mistral.ai/news/codestral-2501},
  note         = {Large Language Model.}
}

@misc{salewski2023incontextimpersonationrevealslarge,
      title={{In-Context Impersonation Reveals Large Language Models' Strengths and Biases}}, 
      author={Leonard Salewski and Stephan Alaniz and Isabel Rio-Torto and Eric Schulz and Zeynep Akata},
      year={2023},
      eprint={2305.14930},
      archivePrefix={arXiv},
      primaryClass={cs.AI},
      url={https://arxiv.org/abs/2305.14930}, 
}

@misc{xu2024carepromptbiasinvestigating,
      title={{Take Care of Your Prompt Bias! Investigating and Mitigating Prompt Bias in Factual Knowledge Extraction}}, 
      author={Ziyang Xu and Keqin Peng and Liang Ding and Dacheng Tao and Xiliang Lu},
      year={2024},
      eprint={2403.09963},
      archivePrefix={arXiv},
      primaryClass={cs.CL},
      url={https://arxiv.org/abs/2403.09963}, 
}

@misc{ahmed2022fewshottrainingllmsprojectspecific,
      title={{Few-shot training LLMs for project-specific code-summarization}}, 
      author={Toufique Ahmed and Premkumar Devanbu},
      year={2022},
      eprint={2207.04237},
      archivePrefix={arXiv},
      primaryClass={cs.SE},
}

@misc{gao2024retrievalaugmentedgenerationlargelanguage,
      title={{Retrieval-Augmented Generation for Large Language Models: A Survey}}, 
      author={Yunfan Gao and Yun Xiong and Xinyu Gao and Kangxiang Jia and Jinliu Pan and Yuxi Bi and Yi Dai and Jiawei Sun and Meng Wang and Haofen Wang},
      year={2024},
      eprint={2312.10997},
      archivePrefix={arXiv},
      primaryClass={cs.CL}
}

@misc{lewis2021retrievalaugmentedgenerationknowledgeintensivenlp,
      title={{Retrieval-Augmented Generation for Knowledge-Intensive NLP Tasks}}, 
      author={Patrick Lewis and Ethan Perez and Aleksandra Piktus and Fabio Petroni and Vladimir Karpukhin and Naman Goyal and Heinrich Küttler and Mike Lewis and Wen-tau Yih and Tim Rocktäschel and Sebastian Riedel and Douwe Kiela},
      year={2021},
      eprint={2005.11401},
      archivePrefix={arXiv},
      primaryClass={cs.CL} 
}

@misc{zhang2020bertscoreevaluatingtextgeneration,
      title={{BERTScore: Evaluating Text Generation with BERT}}, 
      author={Tianyi Zhang and Varsha Kishore and Felix Wu and Kilian Q. Weinberger and Yoav Artzi},
      year={2020},
      eprint={1904.09675},
      archivePrefix={arXiv},
      primaryClass={cs.CL},
      url={https://arxiv.org/abs/1904.09675}, 
}

@inproceedings{lin-2004-rouge,
    title="{ROUGE}: A Package for Automatic Evaluation of Summaries",
    author = "Lin, Chin-Yew",
    booktitle="Text Summarization Branches Out",
    month = jul,
    year = "2004",
    address = "Barcelona, Spain",
    publisher = "Association for Computational Linguistics",
    url = "https://aclanthology.org/W04-1013/",
    pages = "74--81"
}

@inproceedings{10.3115/1073083.1073135,
author = {Papineni, Kishore and Roukos, Salim and Ward, Todd and Zhu, Wei-Jing},
title={{BLEU: a method for automatic evaluation of machine translation}},
year = {2002},
publisher = {Association for Computational Linguistics},
address = {USA},
url = {https://doi.org/10.3115/1073083.1073135},
doi = {10.3115/1073083.1073135},
booktitle={{Proceedings of the 40th Annual Meeting on Association for Computational Linguistics}},
pages = {311–318},
numpages = {8},
location = {Philadelphia, Pennsylvania},
series = {ACL '02}
}

@misc{fu2023gptscoreevaluatedesire,
      title={{GPTScore: Evaluate as You Desire}}, 
      author={Jinlan Fu and See-Kiong Ng and Zhengbao Jiang and Pengfei Liu},
      year={2023},
      eprint={2302.04166},
      archivePrefix={arXiv},
      primaryClass={cs.CL},
}

@misc{fan2024evaluatingqualitycodecomments,
      title={{Evaluating the Quality of Code Comments Generated by Large Language Models for Novice Programmers}}, 
      author={Aysa Xuemo Fan and Arun Balajiee Lekshmi Narayanan and Mohammad Hassany and Jiaze Ke},
      year={2024},
      eprint={2409.14368},
      archivePrefix={arXiv},
      primaryClass={cs.SE},
}

@misc{fried2023incodergenerativemodelcode,
      title={{InCoder: A Generative Model for Code Infilling and Synthesis}}, 
      author={Daniel Fried and Armen Aghajanyan and Jessy Lin and Sida Wang and Eric Wallace and Freda Shi and Ruiqi Zhong and Wen-tau Yih and Luke Zettlemoyer and Mike Lewis},
      year={2023},
      eprint={2204.05999},
      archivePrefix={arXiv},
      primaryClass={cs.SE},
}

@ARTICLE{7997917,
  author={Xia, Xin and Bao, Lingfeng and Lo, David and Xing, Zhenchang and Hassan, Ahmed E. and Li, Shanping},
  journal={IEEE Transactions on Software Engineering}, 
  title={{Measuring Program Comprehension: A Large-Scale Field Study with Professionals}}, 
  year={2018},
  volume={44},
  number={10},
  pages={951-976},
  doi={10.1109/TSE.2017.2734091},
}

@INPROCEEDINGS{981648,
  author={Aggarwal, K.K. and Singh, Y. and Chhabra, J.K.},
  booktitle={{Annual Reliability and Maintainability Symposium. 2002 Proceedings (Cat. No.02CH37318)}}, 
  title={{An integrated measure of software maintainability}}, 
  year={2002},
  volume={},
  number={},
  pages={235-241},
  doi={10.1109/RAMS.2002.981648},
}

@ARTICLE{6171,
  author={Tenny, T.},
  journal={IEEE Transactions on Software Engineering}, 
  title={{Program readability: procedures versus comments}}, 
  year={1988},
  volume={14},
  number={9},
  pages={1271-1279},
  doi={10.1109/32.6171},
}

@misc{shahbazi2021api2comimprovementautomaticallygenerated,
      title={{API2Com: On the Improvement of Automatically Generated Code Comments Using API Documentations}}, 
      author={Ramin Shahbazi and Rishab Sharma and Fatemeh H. Fard},
      year={2021},
      eprint={2103.10668},
      archivePrefix={arXiv},
      primaryClass={cs.SE},
      url={https://arxiv.org/abs/2103.10668}, 
}

@misc{yang2021comformercodecommentgeneration,
      title={{ComFormer: Code Comment Generation via Transformer and Fusion Method-based Hybrid Code Representation}}, 
      author={Guang Yang and Xiang Chen and Jinxin Cao and Shuyuan Xu and Zhanqi Cui and Chi Yu and Ke Liu},
      year={2021},
      eprint={2107.03644},
      archivePrefix={arXiv},
      primaryClass={cs.SE} 
}

@INPROCEEDINGS{4400153,
  author={Fluri, Beat and Wursch, Michael and Gall, Harald C.},
  booktitle={{14th Working Conference on Reverse Engineering (WCRE 2007)}}, 
  title={{Do Code and Comments Co-Evolve? On the Relation between Source Code and Comment Changes}}, 
  year={2007},
  volume={},
  number={},
  pages={70-79},
  doi={10.1109/WCRE.2007.21},
}

@InProceedings{10.1007/978-3-319-19243-7_10,
author="Shmerlin, Yulia
and Hadar, Irit
and Kliger, Doron
and Makabee, Hayim",
editor="Persson, Anne
and Stirna, Janis",
title="To Document or Not to Document? An Exploratory Study on Developers' Motivation to Document Code",
booktitle="Advanced Information Systems Engineering Workshops",
year="2015",
publisher="Springer International Publishing",
address="Cham",
pages="100--106",
isbn="978-3-319-19243-7"
}

@misc{cui2022codeexpexplanatorycodedocument,
      title={{CodeExp: Explanatory Code Document Generation}}, 
      author={Haotian Cui and Chenglong Wang and Junjie Huang and Jeevana Priya Inala and Todd Mytkowicz and Bo Wang and Jianfeng Gao and Nan Duan},
      year={2022},
      eprint={2211.15395},
      archivePrefix={arXiv},
      primaryClass={cs.CL},
}

@misc{venkatkrishna2023docgengeneratingdetailedparameter,
      title={{DocGen: Generating Detailed Parameter Docstrings in Python}}, 
      author={Vatsal Venkatkrishna and Durga Shree Nagabushanam and Emmanuel Iko-Ojo Simon and Melina Vidoni},
      year={2023},
      eprint={2311.06453},
      archivePrefix={arXiv},
      primaryClass={cs.SE},
      url={https://arxiv.org/abs/2311.06453}, 
}

@article{majumdar2022automated,
  title={{Automated evaluation of comments to aid software maintenance}},
  author={Majumdar, Srijoni and Bansal, Ayush and Das, Partha Pratim and Clough, Paul D and Datta, Kausik and Ghosh, Soumya Kanti},
  journal={Journal of Software: Evolution and Process},
  volume={34},
  number={7},
  pages={e2463},
  year={2022},
  publisher={Wiley Online Library}
}

@report{cern1998introdoc,
  title={{Introduction to Software Documentation}},
  author = {Alberto Aimer},
  url = {https://cds.cern.ch/record/383260/files/p135.pdf},
  year = {1998}
}

@article{10.1007/s10664-019-09730-9,
author = {Hu, Xing and Li, Ge and Xia, Xin and Lo, David and Jin, Zhi},
title={{Deep code comment generation with hybrid lexical and syntactical information}},
year = {2020},
issue_date = {May 2020},
publisher = {Kluwer Academic Publishers},
address = {USA},
volume = {25},
number = {3},
issn = {1382-3256},
url = {https://doi.org/10.1007/s10664-019-09730-9},
doi = {10.1007/s10664-019-09730-9},
journal = {Empirical Softw. Engg.},
month = may,
pages = {2179–2217},
numpages = {39}
}

@misc{lu2021codexgluemachinelearningbenchmark,
      title={{CodeXGLUE: A Machine Learning Benchmark Dataset for Code Understanding and Generation}}, 
      author={Shuai Lu and Daya Guo and Shuo Ren and Junjie Huang and Alexey Svyatkovskiy and Ambrosio Blanco and Colin Clement and Dawn Drain and Daxin Jiang and Duyu Tang and Ge Li and Lidong Zhou and Linjun Shou and Long Zhou and Michele Tufano and Ming Gong and Ming Zhou and Nan Duan and Neel Sundaresan and Shao Kun Deng and Shengyu Fu and Shujie Liu},
      year={2021},
      eprint={2102.04664},
      archivePrefix={arXiv},
      primaryClass={cs.SE} 
}

@inproceedings{dong-lapata-2016-language,
    title="Language to Logical Form with Neural Attention",
    author = "Dong, Li  and
      Lapata, Mirella",
    editor = "Erk, Katrin  and
      Smith, Noah A.",
    booktitle="Proceedings of the 54th Annual Meeting of the Association for Computational Linguistics (Volume 1: Long Papers)",
    month = aug,
    year = "2016",
    address = "Berlin, Germany",
    publisher = "Association for Computational Linguistics",
    url = "https://aclanthology.org/P16-1004/",
    doi = "10.18653/v1/P16-1004",
    pages = "33--43"
}

@INPROCEEDINGS{8816757,
  author={Biswas, Sumon and Islam, Md Johirul and Huang, Yijia and Rajan, Hridesh},
  booktitle={{2019 IEEE/ACM 16th International Conference on Mining Software Repositories (MSR)}}, 
  title={{Boa Meets Python: A Boa Dataset of Data Science Software in Python Language}}, 
  year={2019},
  volume={},
  number={},
  pages={577-581},
  doi={10.1109/MSR.2019.00086}
}

@article{Pearson01071900,
author = {Karl Pearson},
title = {{X. On the criterion that a given system of deviations from the probable in the case of a correlated system of variables is such that it can be reasonably supposed to have arisen from random sampling }},
journal = {The London, Edinburgh, and Dublin Philosophical Magazine and Journal of Science},
volume = {50},
number = {302},
pages = {157--175},
year = {1900},
publisher = {Taylor \& Francis},
doi = {10.1080/14786440009463897},
URL = {https://doi.org/10.1080/14786440009463897}
}

@article{chatterjee2023parallelc,
title={{ParallelC-Assist: Productivity Accelerator Suite Based on Dynamic Instrumentation}},
author={Nachiketa Chatterjee and Srijoni Majumdar and Partha Pratim Das and Amlan Chakrabarti},
journal={IEEE Access},
volume={11},
pages={73599--73612},
year={2023},
publisher={IEEE}
}

@article{10.1016/j.jss.2022.111515,
author = {Rani, Pooja and Blasi, Arianna and Stulova, Nataliia and Panichella, Sebastiano and Gorla, Alessandra and Nierstrasz, Oscar},
title = {A decade of code comment quality assessment: A systematic literature review},
year = {2023},
issue_date = {Jan 2023},
publisher = {Elsevier Science Inc.},
address = {USA},
volume = {195},
number = {C},
issn = {0164-1212},
url = {https://doi.org/10.1016/j.jss.2022.111515},
doi = {10.1016/j.jss.2022.111515},
journal = {J. Syst. Softw.},
month = jan,
numpages = {22}
}

@misc{figl2025influencecodecommentsperceived,
      title={The Influence of Code Comments on the Perceived Helpfulness of Stack Overflow Posts}, 
      author={Kathrin Figl and Maria Kirchner and Sebastian Baltes and Michael Felderer},
      year={2025},
      eprint={2508.19610},
      archivePrefix={arXiv},
      primaryClass={cs.SE}, 
}

@INPROCEEDINGS{9610635,
  author={Rani, Pooja and Birrer, Mathias and Panichella, Sebastiano and Ghafari, Mohammad and Nierstrasz, Oscar},
  booktitle={2021 IEEE 21st International Working Conference on Source Code Analysis and Manipulation (SCAM)}, 
  title={What Do Developers Discuss about Code Comments?}, 
  year={2021},
  volume={},
  number={},
  pages={153-164},
  doi={10.1109/SCAM52516.2021.00027}}

@techreport{dart1993case,
  author = {Dart, Susan A. and Christie, Alan M. and Brown, Alan W.},
  title = {A Case Study in Software Maintenance},
  institution = {Software Engineering Institute, Carnegie Mellon University},
  year = {1993},
  number = {CMU/SEI-93-TR-8},
  month = {June},
  address = {Pittsburgh, PA},
  url = {https://www.sei.cmu.edu/documents/1079/1993_005_001_16172.pdf}
}

@misc{rollbar2021,
  author = {Rollbar},
  title = {The 2021 State of Software Code Report},
  year = {2021},
  month = {Feb},
  howpublished = {\url{https://rollbar.com/blog/announcing-the-2021-state-of-software-code-report/}}
}

@misc{linearb2024,
  author = {LinearB},
  title = {Engineering Investment Analysis: Metrics Guide},
  year = {2024},
  howpublished = {\url{https://linearb.io/metrics-guide/}},
  note = {Analysis of 3,000+ teams}
}

@inproceedings{10.1145/1085313.1085331,
author = {de Souza, Sergio Cozzetti B. and Anquetil, Nicolas and de Oliveira, K\'{a}thia M.},
title = {A study of the documentation essential to software maintenance},
year = {2005},
isbn = {1595931759},
publisher = {Association for Computing Machinery},
address = {New York, NY, USA},
url = {https://doi.org/10.1145/1085313.1085331},
doi = {10.1145/1085313.1085331},
booktitle = {Proceedings of the 23rd Annual International Conference on Design of Communication: Documenting \& Designing for Pervasive Information},
pages = {68–75},
numpages = {8},
location = {Coventry, United Kingdom},
series = {SIGDOC '05}
}

@article{DBLP:journals/corr/abs-2005-14165,
  author       = {Tom B. Brown and
                  Benjamin Mann and
                  Nick Ryder and
                  Melanie Subbiah and
                  Jared Kaplan and
                  Prafulla Dhariwal and
                  Arvind Neelakantan and
                  Pranav Shyam and
                  Girish Sastry and
                  Amanda Askell and
                  Sandhini Agarwal and
                  Ariel Herbert{-}Voss and
                  Gretchen Krueger and
                  Tom Henighan and
                  Rewon Child and
                  Aditya Ramesh and
                  Daniel M. Ziegler and
                  Jeffrey Wu and
                  Clemens Winter and
                  Christopher Hesse and
                  Mark Chen and
                  Eric Sigler and
                  Mateusz Litwin and
                  Scott Gray and
                  Benjamin Chess and
                  Jack Clark and
                  Christopher Berner and
                  Sam McCandlish and
                  Alec Radford and
                  Ilya Sutskever and
                  Dario Amodei},
  title        = {Language Models are Few-Shot Learners},
  journal      = {CoRR},
  volume       = {abs/2005.14165},
  year         = {2020},
  eprinttype    = {arXiv},
  eprint       = {2005.14165},
  bibsource    = {dblp computer science bibliography, https://dblp.org}
}

@misc{sun2023abstractsyntaxtreeprogramming,
      title={Abstract Syntax Tree for Programming Language Understanding and Representation: How Far Are We?}, 
      author={Weisong Sun and Chunrong Fang and Yun Miao and Yudu You and Mengzhe Yuan and Yuchen Chen and Quanjun Zhang and An Guo and Xiang Chen and Yang Liu and Zhenyu Chen},
      year={2023},
      eprint={2312.00413},
      archivePrefix={arXiv},
      primaryClass={cs.SE},
      url={https://arxiv.org/abs/2312.00413}, 
}
\end{document}